\documentstyle[prc,aps,twocolumn,epsfig]{revtex}
\begin{document}
\draft
\twocolumn[\hsize\textwidth\columnwidth\hsize\csname
@twocolumnfalse\endcsname

\title{\bf The Bose-Einstein correlation function $C_2(Q)$ from a
Quantum Field Theory point of view}
\author{G.A.Kozlov$^{1}$, O.V.Utyuzh$^{2}$ and G.Wilk$^{2}$}
\address{$^{1}$Bogoliubov Laboratory of Theoretical Physics,
Joint Institute for Nuclear Research, 141980 Dubna,
Moscow Region, Russia}
\address{$^{2}$The Andrzej So\l tan Institute for Nuclear Studies,
Ho\.za 69; 00-689 Warsaw, Poland}
\date{\today}
\maketitle

\begin{abstract}
We show that a recently proposed derivation of Bose-Einstein
correlations (BEC) by means of a specific version of thermal
Quantum Field Theory (QFT), supplemented by operator-field evolution
of the Langevin type, allows for a deeper understanding of the
possible coherent behaviour of the emitting source and a clear
identification of the origin of the observed shape of the BEC
function $C_2(Q)$. Previous conjectures in this matter obtained by
other approaches are confirmed and have received complementary
explanation. 
\end{abstract}

]

In this work we would like to focus attention on two specific
features of Bose-Einstein correlations (BEC) clearly visible when
BEC are presented in the language of some specific (thermal) version
of Quantum Field Theory (QFT) supplemented by the operator-field
evolution of Langevin type proposed recently \cite{Kozlov,Langevin}.
Because the importance of BEC and their present experimental and
theoretical status are widely known and well documented (see, for
example, \cite{BEC} and references therein), we shall not repeat it
here. The features we shall discuss are: $(i)$ how the possible
coherence of the hadronizing system (modelled here by some external
stationary force occurring in the Langevin equations describing
hadronization process \cite{Langevin}) influences the BEC function
$C_2(Q)$ and $(ii)$ what is the true origin of the experimentally
observed $Q$-dependence of the $C_2(Q)$ correlation function in the
approach used here .  

In what concerns the first point we have obtained identical
expression as derived in \cite{Weiner} by means of a quantum
statistical (QS) approach with novel interpretation of the {\it
chaoticity} parameter $p$ introduced there. On the other hand, our
results (and that of \cite{Weiner}) differ from the formula recently
obtained in \cite{ALS}. We shall argue that the origin of this
difference lies in different ways of introducing the concept of
coherence in both approaches. 

In the second point we demonstrate that in order to obtain a
given (experimentally observed) shape of the BEC correlation
function $C_2(Q)$ (i.e., its $Q$-dependence, where
$Q=|k_{\mu}-k'_{\mu}| = \sqrt{(k_{\mu}- k'_{\mu})^2}$) one has to
account somehow for  the {\it finiteness} of the space-time
region of the particle production (i.e., of the hadronizing {\it
source}). In our approach this means the necessity of smearing
out of some generalized functions (delta functions:
$\delta(Q_{\mu}=k_{\mu}-k'_{\mu})$) appearing in the definition of
thermal averages of some operators, which is characteristic feature
of QFT approach used here. The freedom in using different types of
smearing functions to perform such a procedure allows us to account
for all possible different shapes of hadronizing sources apparently
observed by experiment. \cite{FOOTA}. 

Referring to \cite{Kozlov} for details, let us recapitulate here the
main points of our approach. The collision process produces a lot of
particles out of which we select one (we assume for simplicity that
we are dealing only with identical bosons) and describe it by
operator $b(\vec{k},t)$ (the notation is the usual one:
$b(\vec{k},t)$ is an annihilation operator, $\vec{k}$ is
$3$-momentum and $t$ is a real time). The rest of the particles
are then assumed to form a kind of heat bath, which remains in
equilibrium characterized by a temperature $T=1/\beta$ (which will be
one of our parameters). All averages $\langle (\dots) \rangle$ are
therefore thermal averages of the type: $\langle (\dots) \rangle
=Tr\left[(\dots) e^{-\beta H}\right]/Tr\left( e^{-\beta H}\right)$. 
We shall also allow for some external (to the above heat bath)
influence to our system. Therefore we shall represent the operator
$b(\vec{k},t)$ as consisting of a part corresponding to the action
of the heat bath, $a(\vec{k},t)$, and also of a part describing action of
these external factors, $R(\vec{k},t)$:
\begin{equation}
b(\vec{k},t) = a(\vec{k},t) + R(\vec{k},t).\label{eq:apR}
\end{equation}
The time evolution of such a system is then assumed to be given by a
Langevin equation \cite{Langevin}
\begin{equation}
i\partial_t b(\vec{k},t) = F(\vec{k},t) - A(\vec{k},t) + P
\label{eq:Lang}
\end{equation}
(and a similar conjugate equation for $b^{+}(\vec{k},t)$). These
equations are supposed to model all aspects of the hadronization
process. The combination $F(\vec{k},t)-A(\vec{k},t)$ represents the
so called {\it Langevin force} and is therefore responsible for the
internal dynamics of hadronization in the following manner: $A$ is
related to stochastic dissipative forces and is given by
\cite{Langevin,Kozlov} 
\begin{equation}
A(\vec{k},t) = \int^{+\infty}_{-\infty}\! d\tau K(\vec{k},t-\tau)
b(\vec{k},\tau), \label{eq:AL}
\end{equation}
with the operator $K(\vec{k},t)$ being a random evolution field
operator describing the random noise and satisfying the usual
correlation-fluctuation relation for the Gaussian noise
\cite{FOOTB}.
The operator $F(\vec{k},t)$ describes the influence of heat bath,
\begin{equation}
F(\vec{k},t) =
\int^{+\infty}_{-\infty}\!\frac{d\omega}{2\pi}\psi(k_{\mu})\hat{c}(k_{\mu})
e^{-i\omega t} . \label{eq:FL}
\end{equation}
Our heat bath is represented by an ensemble of damped oscillators,
each described by operator $\hat{c}(k_{\mu})$ such
that $\left[\hat{c}(k_{\mu}),\hat{c}^{+}(k'_{\mu})\right] =
\delta^4(k_{\mu}-k'_{\mu})$, and characterized by some function
$\psi(k_{\mu})$ \cite{FOOTI}.
Finally, the constant term $P$ (representing {\it external source}
term in Langevin equation) denotes the possible influence of
some external force (assumed here to be constant in time). This force
would result, for example, in a strong ordering of phases, leading 
therefore to the coherence effect in the sense discussed in \cite{Weiner}.
Out of many details (for which we refer to \cite{Kozlov}) what is important
in our case is the fact that the $2$-particle correlation function
for like-charge particles, is defined as $(k_{\mu} =
(\omega=k^{0},k_j)$):
\begin{eqnarray}
C_2(Q) &=&  \xi(N) \cdot \frac{\tilde{f}(k_{\mu},k'_{\mu})}
{\tilde{f}(k_{\mu})\cdot\tilde{f}(k'_{\mu})}\nonumber\\
&=& \xi(N) \cdot \left[1\, +\, D(k_{\mu},k'_{\mu})\right] ,
\label{eq:C2}
\end{eqnarray}
where $\tilde{f}(k_{\mu},k'_{\mu}) = \langle
\tilde{b}^{+}(k_{\mu})\tilde{b}^{+}(k'_{\mu})
\tilde{b}(k_{\mu})\tilde{b}(k'_{\mu})\rangle$ and
$\tilde{f}(k_{\mu}) = \langle
\tilde{b}^{+}(k_{\mu})\tilde{b}(k_{\mu})\rangle$ are the
corresponding thermal statistical averages (in which temperature $T$
enters as a parameter) with $\tilde{b}(k_{\mu}) = \tilde{a}(k_{\mu})
+ \tilde{R}(k_{\mu})$
being the corresponding Fourier transformed stationary solution
of eq. (\ref{eq:Lang}). As shown in \cite{Kozlov} (notice that
operators $\tilde{R}(k_{\mu})$ by definition commute with themselves
and with any other operator considered here):
\begin{eqnarray}
\tilde{f}(k_{\mu},k'_{\mu}) &=& \tilde{f}(k_{\mu})\cdot\tilde{f}(k'_{\mu}) +\nonumber\\
&+& \langle\tilde{a}^{+}(k_{\mu})\tilde{a}(k'_{\mu})\rangle
\langle\tilde{a}^{+}(k'_{\mu})\tilde{a}(k_{\mu})\rangle  + \nonumber\\
&+&\langle\tilde{a}^{+}(k_{\mu})\tilde{a}(k'_{\mu})\rangle \tilde{R}^{+}(k'_{\mu})\tilde{R}(k_{\mu}) +\nonumber\\
&+&\langle\tilde{a}^{+}(k'_{\mu})\tilde{a}(k_{\mu})\rangle \tilde{R}^{+}(k_{\mu})\tilde{R}(k'_{\mu}) ,\label{eq:ff}\\
\tilde{f}(k_{\mu}) &=& \langle \tilde{a}^+(k_{\mu})\tilde{a}(k_{\mu})\rangle\, +\,
|\tilde{R}(k_{\mu})|^2 . \label{eq:f}
\end{eqnarray}
This defines $D(k_{\mu},k'_{\mu}) = \tilde{f}(k_{\mu},k'_{\mu})/
[\tilde{f}(k_{\mu})\cdot\tilde{f}(k'_{\mu})] - 1$ in (\ref{eq:C2}) in 
terms of the operators $\tilde{a}(k_{\mu})$ and $\tilde{R}(k_{\mu})$, 
which in our case are equal to:
\begin{equation}
\tilde{a}(k_{\mu}) =
\frac{\tilde{F}(k_{\mu})}{\tilde{K}(k_{\mu}) - \omega}\quad {\rm and}\quad
\tilde{R}(k_{\mu}) =
\frac{P}{\tilde{K}(k_{\mu}) - \omega} . \label{eq:aR}
\end{equation}
The multiplicity $N$ depending factor $\xi$ is in our case equal to $\xi
(N) = \langle N\rangle^2/\langle N(N-1)\rangle$. This means therefore
that the correlation function $C_2(Q)$, as defined
by eq. (\ref{eq:C2}), is essentially given in terms of $P$ and the
two following thermal averages for
the $F(\vec{k},t)$ operators:
\begin{eqnarray}
\langle F^{+}(\vec{k},t)F(\vec{k}',t')\rangle &=&
\delta^3(\vec{k}-\vec{k}')\cdot\nonumber\\
&\cdot&\int \frac{d\omega}{2\pi}\,
               \left|\psi\right|^2\, n(\omega)e^{+i\omega(t-t')},
               \label{eq:theavc}\\
\langle F(\vec{k},t)F^{+}(\vec{k}',t')\rangle &=&
\delta^3(\vec{k}-\vec{k}')\cdot \nonumber\\
&\cdot&\int \frac{d\omega}{2\pi}\,
               \left|\psi\right|^2\, [1 + n(\omega)] e^{-i\omega(t-t')}\nonumber\\
\end{eqnarray}
where $n(\omega) = \left\{\exp \left[(\omega - \mu)\beta\right] -
1\right\}^{-1}$ is the number of (by assumption - only bosonic in our
case) damped oscillators of energy $\omega$ in our reservoir
characterized by parameters $\mu$ (chemical potential) and inverse
temperature $\beta=1/T$ (both being free parameters)\cite{FOOTII}.
Notice that with only delta functions present
in (\ref{eq:theavc}) one would have a situation in which our hadronizing
system would be described by some kind of {\it white noise} only. The
integrals multiplying these delta functions and depending on $(a)$ momentum
characteristic of our heat bath $\psi(k_{\mu})$ (representing in our case,
by definition, the hadronizing system) and $(b)$ assumed bosonic statistics
of produced secondaries resulting in factors $n(\omega)$ and $1+n(\omega)$,
respectively, bring the description of our system closer to
reality.

It should be stressed at this point that, contrary to the majority of
discussions of BEC \cite{BEC,Weiner}, we are working here directly in
phase space \cite{Takagi}, so far no space-time considerations
were used. It is easy to realize now that the existence of BEC, i.e.,
the fact that $C_2(Q) >1$, is strictly connected with nonzero values
of the thermal averages (\ref{eq:theavc}). However, in the form
presented there, they differ from zero {\it only at one point},
namely for $Q=0$ (i.e., for $k_{\mu} = k'_{\mu}$). Actually, this is
the price one pays for the QFT assumptions tacitly made here, namely
for the {\it infinite} spatial extension and for the {\it uniformity}
of our reservoir. But we know from the experiment \cite{BEC} that
$C_2(Q)$ reaches its maximum at $Q=0$ and falls down towards its
asymptotic value of $C_2 = 1$ at large of $Q$ (actually already at $Q
\sim 1$ GeV/c). To reproduce the same behaviour by means of our
approach here, one has to replace delta functions in eq.
(\ref{eq:theavc}) by functions with supports larger than limited to a
one point only. This means that such functions should not be infinite
at $Q_{\mu} = k_{\mu}-k'_{\mu} =0$ but remain more or less sharply
peaked at this point, otherwise remaining finite and falling to zero
at small, but finite, values of $|Q_{\mu}|$ (actually the same as
those at which $C_2(Q)$ reaches unity): 
\begin{equation}
\delta(k_{\mu} - k'_{\mu})\, \Longrightarrow\, \Omega_0\cdot
\sqrt{\Omega(q=Q\cdot r)}.
\label{eq:Om}
\end{equation}
Here $\Omega_0$ has the same dimension as the $\delta$ function
(actually, it is nothing else but $4$-dimensional volume restricting
the space-time region of particle production) and $\Omega(q)$ is a
dimensionless smearing function which contains the $q$-dependence we
shall be interested in here.  In this way we are tacitly introducing
a new parameter, $r_{\mu}$, a $4$-vector such that
$\sqrt{(r_{\mu})^2}$ has dimension of length and which makes the
product $Q\cdot r = Q_{\mu} r_{\mu}= q$ dimensionless. This defines the
region of {\it nonvanishing} density of oscillators $\hat{c}$, which
we shall {\it identify} with the space-time extensions of the
hadronizing source. The expression (\ref{eq:Om}) has to be understood
in a symbolic sense, i.e., that $\Omega(Q\cdot r)$ is a function
which in the limit of $r\rightarrow \infty$ becomes {\it strictly} a
$\delta$ function. Making such replacement in eq. (\ref{eq:theavc})
one must also decide how to accordingly adjust $n(\omega)$ occurring
there because now, in general, $\omega \neq \omega'$. In what follows
we shall simply replace $n(\omega) \rightarrow n(\bar{\omega})$ with
$\bar{\omega} = (\omega + \omega')/2$ (which, for classical particles
would mean that $n(\omega) \rightarrow \sqrt{n(\omega)n(\omega')}$).

In such way $r$ becomes new (and from the phenomenological point of
view also the most important) parameter entering here together with
the whole function $\Omega(Q\cdot r)$, to be deduced from comparison
with experimental data \cite{Z}. With such a replacement one now has
\begin{equation}
D(k_{\mu},k'_{\mu}) =
\frac{\sqrt{\tilde{\Omega}(q)}}{(1+\alpha)(1+\alpha')}
\cdot \left[ \sqrt{\tilde{\Omega}(q)} + 2\sqrt{\alpha \alpha'}
\right]  \label{eq:res}
\end{equation}
where
\begin{equation}
\tilde{\Omega}(q) = \gamma \cdot \Omega(q),~~
\gamma = \frac{n^2(\bar{\omega})}{n(\omega)n(\omega')},~~
\alpha \propto \frac{P^2}{|\psi(k_{\mu})|^2 n(\omega)},
\label{eq:lambda}
\end{equation}
with $n(\omega)$ the same as defined above. The parameter $\alpha$ is
another very important parameter, which summarizes our knowledge of
other than space-time characteristics of the hadronizing source
(given by $\Omega(q)$ introduced above). In particular it contains
the external static force $P$ present in the evolution equation
(\ref{eq:Lang}). It is combined (in multiplicative way) with information
on the momentum
dependence of the reservoir (via $|\psi(k_{\mu})|^2$) and on the
single particle distributions of the produced particles (via
$n(\omega = \mu_T \cosh y)$ where $\mu_T$ and $y$ are, respectively,
the transverse mass and rapidity). Notice that $\alpha > 0$
only when $P \neq 0$. Actually, for $\alpha = 0$ one has
\begin{equation}
1 < C_2(Q) < 1 + \gamma \Omega(Q\cdot r) ,
\label{eq:limits}
\end{equation}
i.e., it is contained between limits corresponding to very large
(lower limit) and very small (upper limit) values of $P$.
Because of this $\alpha$ plays the role of the {\it coherence}
parameter \cite{BEC,Weiner}. For $\gamma \simeq 1$, neglecting
the possible energy-momentum dependence of $\alpha$ and assuming that
$\alpha' = \alpha$ one gets the expression
\begin{equation}
C_2(Q) = 1 + \frac{2\alpha}{(1 + \alpha)^2}\cdot
\sqrt{\Omega(q)}\, +\, \frac{1}{(1+\alpha)^2}\cdot
\Omega(q) , \label{eq:C2final}
\end{equation}
which is formally {\it identical} with what has been obtained in
\cite{Weiner} by means of QS approach. It has precisely the same
form, consisting two $Q-$dependent terms containing the information
on the shape of the source, one being the square of the other, each
multiplied by some combination of the {\it chaoticity} parameter $p =
1/(1+\alpha)$ (however, in \cite{Weiner} $p$ is defined as the ratio
of the mean multiplicity of particles produced by the so called {\it
chaotic} component of the source to the mean total multiplicity, 
$p=\langle N_{ch}\rangle/ \langle N\rangle$).
In fact, because in general $\alpha \neq \alpha'$
(due to the fact that $\omega \neq \omega'$ and therefore the number
of states, identified here with the number of particles with given
energy, $n(\omega)$, are also different) one should rather use the
general form (\ref{eq:C2}) for $C_2$ with details given by
(\ref{eq:res}) and (\ref{eq:lambda}) and with $\alpha$ depending
on such characteristics of the production process as temperature $T$ and
chemical potential $\mu$ occurring in definition of $n(\omega)$.

Notice that eq. (\ref{eq:C2final}) differs from the usual
empirical parameterization of $C_2(Q)$ \cite{BEC},
\begin{equation}
C_2(Q) = 1 + \lambda\cdot \Omega(Q\cdot r), \label{eq:usual}
\end{equation}
with $0< \lambda <1$ being a free parameter adjusting the observed
value of $C_2(Q=0)$, which is customary called "incoherence", and with
$\Omega(Q\cdot r)$ represented usually as Gaussian. Recently eq.
(\ref{eq:usual}) has found strong theoretical support expressed in
great detail in \cite{ALS}. The natural question arises: which of the
two formulas presented here is correct? The answer is: both are right
in their own way. This is because each of them is based on different
ways of defining coherence of the source. In \cite{ALS} one uses the
notion of coherently and chaotically produced particles or, in other
words, one divides hadronizing source into coherent and chaotic
subsources. In \cite{Weiner} one introduces instead the notion of
partially coherent fields representing produced particles, i.e., one
has only one source, which produces partially coherent fields. Our
approach is similar as we describe our particle by operator
$b(\vec{k},t)$, which consists of two parts, cf. eq. (\ref{eq:apR}),
one of which depends on the external static force $P$. The action of
this force is to {\it order phases} of particles in our source
(represented by the heat bath). The strength of this ordering depends
on the value of the external force $P$. In any case, for $P \ne 0$,
it demonstrates itself as a {\it partial coherence} \cite{FOOT-1}.

Let us return to the problem of $Q$-dependence of BEC. One more
remark is in order here. The problem with the
$\delta(k_{\mu}-k'_{\mu})$ function encountered in two particle
distributions does not exist in the single particle distributions, 
which are in our case given by eq. (\ref{eq:f}) and which can be written as
$\tilde{f}(k_{\mu}) \propto
\langle \tilde{a}^+(k_{\mu})\tilde{a}(k_{\mu})\rangle\, +\,
|\tilde{R}(k_{\mu})|^2 \, \sim (1+\alpha)\langle
\tilde{a}^+(k_{\mu})\tilde{a}(k_{\mu})\rangle$ \cite{FOOT3}.
To be more precise
\begin{equation}
\tilde{f}(k_{\mu})\, = \, (1+\alpha) \cdot \Xi(k_{\mu},k_{\mu}) ,
 \label{eq:Single}
\end{equation}
where $\Xi(k_{\mu},k_{\mu})$ is one-particle distribution function
for the "free" (undistorted) operator $\tilde{a}(k_{\mu})$ equal to
\begin{equation}
\Xi(k_{\mu},k_{\mu})\, =\, \Omega_0 \cdot \left| \frac{\psi(k_{\mu})}
                           {\tilde{K}(k_{\mu})-\omega}\right| ^2
n(\omega) . \label{eq:singlea}
\end{equation}
Notice that the actual shape of $\tilde{f}(k_{\mu})$ is dictated both
by $n(\omega) = n(\omega;T,\mu)$ (calculated for fixed temperature
$T$ and chemical potential $\mu$ at energy $\hat{\omega}$ as given by
the Fourier transform of random field operator $\tilde{K}$ and by
shape of the reservoir in the momentum space provided by
$\psi(k_{\mu})$) and by external force $P$ in parameter $\alpha$.
They are both unknown, but because these details do not enter the BEC
function $C_2(Q)$, we shall not pursue this problem further. What is
important for us at the moment is that both the coherent and the
incoherent part of the source have the same energy-momentum
dependence (whereas in other approaches mentioned here they were
usually assumed to be different). On the other hand it is clear from
(\ref{eq:Single}) that $\langle N\rangle = \langle N_{ch}\rangle +
\langle N_{coh}\rangle$ (where $\langle N_{ch}\rangle$ and $\langle
N_{coh}\rangle$ denote multiplicities of particles produced
chaotically and coherently, respectively) therefore justifying
definition of chaoticity $p$ mentioned above.

\begin{figure}
\noindent
\centerline{\epsfig{file=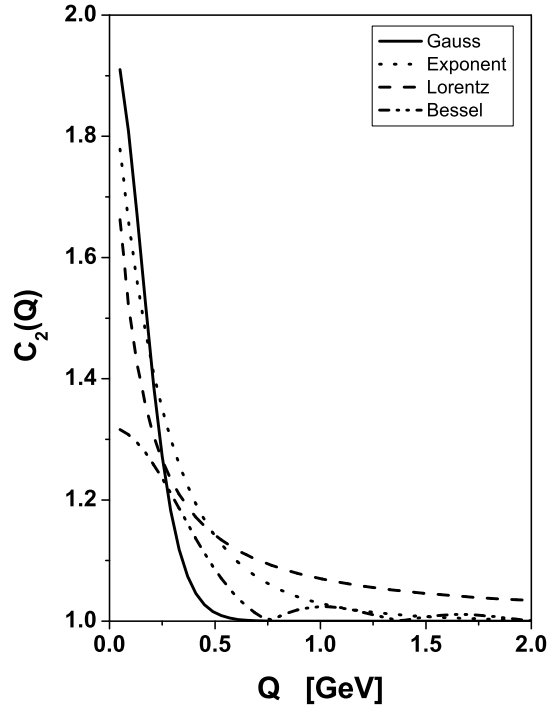, width=72mm}}
\caption{Shapes of $C_2(Q)$ as given by eq. (\protect\ref{eq:C2final})
for different choices of smearing functions $\Omega(q)$ (cf. text for
details).}
\label{fig:plot1}
\end{figure}

For an illustration we plotted in Fig. \ref{fig:plot1} the correlation
function $C_2(Q)$ as given by eq. (\ref{eq:C2final}) for different
choices of $\Omega(q)$ corresponding to different hadronizing sources
discussed in the literature \cite{Sources,Podg,Pratt} (here $r =
|r_{\mu}| = \sqrt{r_{\mu}r_{\mu}}$):
\begin{itemize}
\item Gaussian: $\Omega(q) = \exp \left( - Q^2 r^2\right)$;
\item exponential: $\Omega(q) = \exp ( - Q r)$;
\item Lorentzian: $\Omega(q) = 1/\left(1 + Q r\right)^2$;
\item given by Bessel function \cite{Podg}:\\
  $\Omega(q) = \left[J_1(Q r)/(Q r)\right]^2$ .
\end{itemize}
All curves are drawn for the same values of the size parameter $r=1$
fm and assuming for simplicity constant and equal values of
$\alpha$ and $\alpha'$ parameters (i.e., using simplified eq.
(\ref{eq:C2final})), which have been put equal $\alpha =0.2$ here,
just for illustrational purpose (it corresponds to $p=0.8$ in
\cite{Weiner}). Fig. 2 shows in detail (using Gaussian shape of
$\Omega (q)$ function) the dependence of $C_2(Q)$ on different values of
$\alpha = 0,~0.25,~1,~4$ (again, used in the same approximate way as
before and corresponding to $p=1.,~0.8,~0.5,~0.2$) and compare it to
the case when the second term in eq. (\ref{eq:C2final}) is neglected,
as is the case in majority of phenomenological fits to data.

\begin{figure}
\noindent
\centerline{\epsfig{file=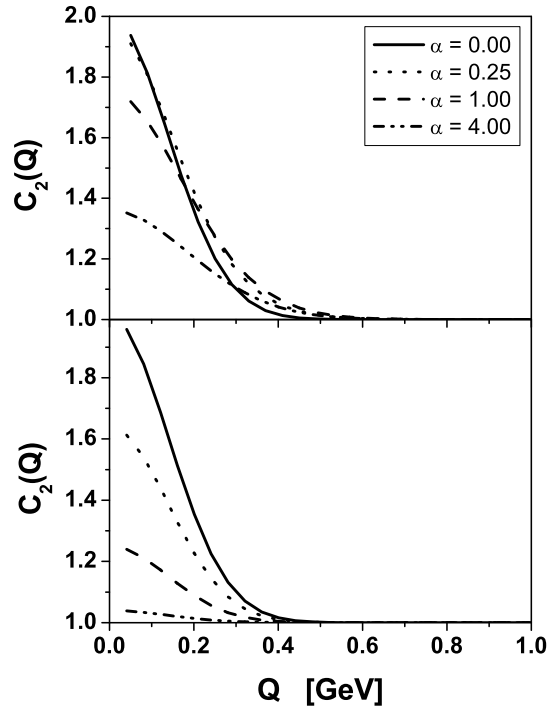, width=72mm}}
\caption{Shapes of $C_2(Q)$ as given by eq. (\protect\ref{eq:C2final}) -
upper panel and for the truncated version of (\protect\ref{eq:C2final})
(without the middle term) - lower panel. Gaussian shape of $\Omega(q)$ was
used in both cases.}
\label{fig:plot2}
\end{figure}

To summarize: using a specific version of QFT supplemented by
Langevin evolution equation (\ref{eq:Lang}) to describe hadronization
process \cite{Kozlov,Langevin} we have derived the usual BEC
correlation function in the form explicitly showing the origin of
both the so called coherence (and how it influences the structure of
BEC) and the $Q$-dependence of BEC represented by correlation
function $C_2(Q)$. The dynamical source of coherence is identified in
our case with the existence of a constant external term $P$ in the
Langevin equation. Its influence turns out to be identical with the
one obtained before in the QS approach \cite{Weiner} and is described
by eq. (\ref{eq:res}). Its action is to order phases of the produced
secondaries. Therefore for $P\rightarrow \infty$ we have all phases
aligned in the same way and $C_2(Q) =1$. This is because both here
and in \cite{Weiner} the coherence has already been introduced on the
level of a hadronizing source, as property of fields (in \cite{Weiner})
or operators describing produced particles. Dividing instead the
hadronizing source itself into coherent and chaotic subsources leads
to results obtained in \cite{ALS} and given by eq. (\ref{eq:usual}).
The controversy between results given by \cite{Weiner} and \cite{ALS}
is therefore explained: both approaches are right, one should only
remember that they use different descriptions of the notion of
coherence. It is therefore up to the experiment to decide which
proposition is followed by nature: the simpler formula
(\ref{eq:usual}) or rather the more involved (\ref{eq:C2}) together
with (\ref{eq:res}). From Fig. \ref{fig:plot2} one can see that
differences between both forms are clearly visible, especially for
larger values of coherence $\alpha$, i.e., for lower chaoticity
parameter $p$. 


From our presentation it is also clear that the form of $C_2$
reflects distributions of the space-time separation between the 
two observed particles rather than the distribution of their separate
production points \cite{Zajc} (i.e., it is Fourier transform of
two-particle density profile of the hadronizing source,
$\rho(r_1,r_2)=\rho(r_1-r_2)$, without approximating it by the
product of single-particle densities, as in \cite{BEC}).\\

Finally, we would like to stress that our discussion is so far
limited to only a single type of secondaries being produced. It is
also aimed at a description of hadronization understood as kinetic
freeze-out in some more detailed approaches. So far we were not
interested in the other (highly model dependent) details of the
particle production process. This is enough to obtain our general
goals, i.e., to explain the possible dynamical origin of coherence in
BEC and the origin of the specific shape of the correlation $C_2(Q)$
functions as seen from the QFT perspective. Actually, our source of
coherence should be regarded as being only one possibility
\cite{BEC,Weiner}, the others were discussed in detail in \cite{ALS}.
It is then plausible that in general description of the BEC effect
they should be somehow combined, especially if experimental data
would indicate such necessity. But to do so our approach should first
be generalized to allowing, as is the case in \cite{ALS}, for production
of different types of secondaries and allow also for resonance production
and final state interactions (both of strong and Coulomb origin). This is,
however, outside the scope of the present paper. \\

We would like to acknowledge support obtained from the
Bogolyubov-Infeld program in JINR and partial support of the Polish
State Committee for Scientific Research (KBN), grants 
621/E-78/SPUB/CERN/P-03/DZ4/99 and 2P03B05724.


\begin{references}
\bibitem{FOOTA} Actually, careful inspection of all previous
                approaches to BEC using QFT (cf., for example,
                \cite{RW}) shows that this was always the procedure used,
                though never expressed so explicitly as is done here.
\bibitem{FOOTB} This means that one demands that
                $\langle K^{+}(\vec{k},t)K(\vec{k}',t)\rangle =
                2\sqrt{\pi \rho}\kappa \delta(\vec{k}-\vec{k}')$
                where $\kappa$ and $\rho$ are parameters defining the
                effect on the particle evolution in thermal
                environment caused by the (assumed here) Gaussian
                noise \cite{Langevin}.
\bibitem{FOOTI} The only condition function $\psi(k_{\mu})$ is
                subjected to is a kind of normalization involving
                also dissipative forces represented by Fourier
                transformed operator $\tilde{K}(k_{\mu})$, namely:
                $\int^{+\infty}_{-\infty}\frac{d\omega}{2\pi}
                \left[\frac{\psi(k_{\mu})}
                {\tilde{K}(k_{\mu})-\omega}\right]^2= 1$.
\bibitem{FOOTII} The origin of the two parameters occurring at this point,
                 the temperature $\beta =1/T$ and chemical potential $\mu$
                 is the Kubo-Martin-Schwinger condition that
                 $\langle a(\vec{k}',t')a^{+}(\vec{k},t)\rangle =
                 \langle a^{+}(\vec{k},t)a({k}',t-i\beta)\rangle
                 \cdot \exp(-\beta \mu)$, see \cite{Kozlov,Langevin}
                 This form of averages reflects the corresponding
                 averages for the $\hat{c}(k_{\mu})$ operators, namely
                 that $\langle \hat{c}^{+}(k_{\mu})\hat{c}(k'_{\mu})\rangle
                 = \delta^4(k_{\mu} - k'_{\mu})\cdot n(\omega)$ and
                 $\langle \hat{c}(k_{\mu})\hat{c}^{+}(k'_{\mu})\rangle =
                 \delta^4(k_{\mu} - k'_{\mu})\cdot [1 + n(\omega)]$.
\bibitem{FOOT3} Notice that it is normalized to the mean
                multiplicity, i.e., $\int d^4k\, \tilde{f}(k_{\mu})\,
                = \, \langle N\rangle$.
\end{references}
\end{document}